\documentclass{article}
\def\deg{{$^{\circ}$}}
\def\bd17{BD +17\deg 3248}
\usepackage{latexsym}
\usepackage{longtable}

\begin{document}
\begin{center}
IMPROVED LABORATORY TRANSITION PROBABILITIES FOR Hf \textsc{ii} AND HAFNIUM 
ABUNDANCES IN THE SUN AND 10 METAL-POOR STARS
\end{center}

\begin{center}
(short title: Hf Transition Probabilities and Abundances)
\end{center}

\begin{center}
J. E. Lawler, E. A. Den Hartog, Z. E. Labby
\end{center}

\begin{center}
Department of Physics, University of Wisconsin, Madison, WI 53706;
\end{center}

\begin{center}
jelawler@wisc.edu, eadenhar@wisc.edu, zelabby@wisc.edu
\end{center}

\begin{center}
C. Sneden
\end{center}

\begin{center}
Department of Astronomy and McDonald Observatory, University of Texas, 
Austin, TX 78712; chris@verdi.as.utexas.edu
\end{center}

\begin{center}
J. J. Cowan
\end{center}

\begin{center}
Homer L. Dodge Department of Physics and Astronomy, University of Oklahoma, 
Norman, OK 73019; cowan@nhn.ou.edu
\end{center}

\begin{center}
and I. I. Ivans
\end{center}

\begin{center}
The Observatories of the Carnegie Institution of Washington, 813 Santa 
Barbara St., Pasadena, CA 91101 {\&} Princeton University Observatory, 
Peyton Hall, Princeton, NJ 08544; iii@ociw.edu
\end{center}

\begin{center}
ABSTRACT
\end{center}
Radiative lifetimes from laser-induced fluorescence measurements, accurate 
to $\sim \pm $5{\%}, are reported for 41 odd-parity levels of Hf 
\textsc{ii}. The lifetimes are combined with branching fractions measured 
using Fourier transform spectrometry to determine transition probabilities 
for 150 lines of Hf \textsc{ii}. Approximately half of these new transition 
probabilities overlap with recent independent measurements using a similar 
approach. The two sets of measurements are found to be in good agreement for 
measurements in common. Our new laboratory data are applied to refine the 
hafnium photospheric solar abundance and to determine hafnium abundances in 
10 metal-poor giant stars with enhanced $r-$process abundances. For the Sun we 
derive log $\varepsilon $(Hf) = 0.88 $\pm $ 0.08 from four lines; the 
uncertainty is dominated by the weakness of the lines and their blending by 
other spectral features. Within the uncertainties of our analysis, the 
$r$-process-rich stars possess constant Hf/La and Hf/Eu abundance ratios, log 
$\varepsilon $(Hf/La) = -0.13 $\pm $ 0.02 ($\sigma $ = 0.06) and log 
$\varepsilon $(Hf/Eu) = +0.04 $\pm $ 0.02 ($\sigma $ = 0.06). The observed 
average stellar abundance ratio of Hf/Eu and La/Eu is larger than previous 
estimates of the solar system $r-$process-only value, suggesting a somewhat 
larger contribution from the $r-$process to the production of Hf and La. The 
newly determined Hf values could be employed as part of the chronometer 
pair, Th/Hf, to determine radioactive stellar ages.

\smallskip
Subject headings: atomic data --- Sun: abundances --- stars: (HD 74462, HD 
115444, HD122956, HD 165195, HD 175305, HD 186478, HD 221170, \bd17, CS 
22892-052, CS 31082-001), nucleosynthesis, abundances --- Galaxy: evolution

\newpage 
\begin{center}
1. INTRODUCTION
\end{center}

Studies of heavy element nucleosynthesis have advanced rapidly in recent 
years due to enormous improvements in observational capabilities and 
significant improvements in basic atomic spectroscopic data. 
High-resolution, high signal-to-noise (S/N) spectra on a variety of targets 
from very large ground-based telescopes and the Hubble Space Telescope are 
now available. Observations of old metal-poor Galactic halo stars are 
central to many of these studies because such stars provide a fossil record 
of the chemical make-up of our Galaxy when it, and the Universe, were very 
young (e.g., Gratton {\&} Sneden 1994; McWilliam et al. 1995; Cowan et al. 
1995; Sneden et al. 1996, Ryan et al. 1996, Cayrel et al. 2004). Abundance 
determinations of heavy $n$(eutron)-capture elements in very metal-poor stars 
are being improved and are steadily yielding new insights on the roles of 
the $r$(apid)- and $s$(low)-processes in the initial burst of Galactic 
nucleosynthesis (e.g. Simmerer et al. 2004, Ivans et al. 2006, Cowan {\&} 
Sneden 2006). Recent theoretical and observational advances have been made 
in metal-poor nucleocosmochronometry studies. The detection of a second 
radioactive element, U, in the spectra of a halo star provides an important 
constraint on age determinations using Th (Cayrel et al. 2001). The results 
of these ongoing studies are reshaping our understanding of the chemical 
evolution of the Galaxy.

This paper reports new results for Hf. The choice of a stable reference 
element for nucleocosmochronometry is a crucial part of an age 
determination. It is essential that the radioactive element, Th (Z = 90) or 
U (Z = 92), and stable reference element(s) were synthesized in the same 
event. Confidence that the radioactive element and reference element have 
the same origin increases if the elements have a similar high Z. Accurate, 
multiple line abundance determinations are difficult for the heaviest 
elements (Z $\sim $ 90), thus there is some compromise required in the 
choice of reference element. We propose that Hf (Z = 72) has good potential 
as a reference element in nucleocosmochronometry. Improved laboratory data, 
especially atomic transition probabilities, are essential for using Hf as a 
reference element in nucleocosmochronometry.

The new laboratory measurements reported in {\S} 2 (radiative lifetimes) and 
{\S} 3 (branching fractions and transition probabilities) of this paper were 
completed shortly before we learned of an overlapping study by Lundqvist et 
al. (2006). The partial overlap of the two sets of measurements provides an 
opportunity to assess systematic uncertainties in modern atomic transition 
probability determinations based on combining radiative lifetimes from laser 
induced fluorescence measurements with branching fractions from Fourier 
transform spectrometry. The good agreement that we find in the comparison of 
the overlapping measurements provides reassurance that estimates of 
systematic uncertainties in modern transition probability measurements are 
reliable.

We apply the new laboratory data to solar and stellar Hf abundances in {\S} 
4. After discussing Hf \textsc{ii} line selection, in {\S} 4.2 we refine 
earlier determinations of the Solar photospheric hafnium abundance, and {\S} 
4.3 we determine the abundance of hafnium in 10 metal-poor giant stars with 
enhanced $r-$process abundances, finding essentially constant Hf/La and Hf/Eu 
abundance ratios. These abundances are discussed in {\S} 5 in the context of 
$r$-process nucleosynthesis.

\begin{center}
2. RADIATIVE LIFETIME MEASUREMENTS
\end{center}

Radiative lifetimes of 41 odd-parity levels of Hf \textsc{ii} have been 
measured using time-resolved laser-induced fluorescence (LIF) on an atom/ion 
beam. Only an overview of the experimental method is given here, since the 
apparatus and technique have been described in many previous publications on 
other species. The reader is referred to recent work in Eu I, II, and III 
(Den Hartog et al. 2002) for a more detailed description.

A hollow cathode discharge sputter source is used to produce a slow ($\sim 
$5$\times $10$^{4}$ cm/s), weakly collimated beam of Hf atoms and ions. A 
pulsed argon discharge, operating at $\sim $0.4 torr with 10 $\mu $s 
duration, 10 A pulses, is used to sputter the hafnium foil which lines the 
hollow cathode. The hollow cathode is closed on one end except for a 1 mm 
hole, through which the hafnium atoms and ions flow into a low pressure 
(10$^{-4}$ torr) scattering chamber. This beam is intersected at right 
angles by a nitrogen laser-pumped dye laser beam 1 cm below the cathode 
bottom. The laser is tunable over the range 3610 to 7200 {\AA} using a wide 
array of commercially available dyes. This range is extended down to 2050 
{\AA} with the use of frequency doubling crystals. The laser is pulsed at 
$\sim $30 Hz repetition rate with a $\sim $3 ns pulse duration, and has a 
$\sim $0.2 cm$^{-1}$ bandwidth. The laser is used to selectively excite the 
level to be studied, eliminating the possibility of cascade radiation from 
higher-lying levels. 

Fluorescence is collected in a direction mutually orthogonal to the laser 
and atomic/ionic beams using a pair of fused-silica lenses that form an f/1 
optical system, and detected with a RCA 1P28A photomultiplier tube (PMT). 
Optical filters, either broadband colored glass filters or narrowband 
multi-layer dielectric filters, can be inserted between the two lenses to 
cut down on scattered laser light and to block cascade radiation from lower 
levels. The signal from the PMT is recorded and averaged over 640 shots 
using a Tektronix SCD1000 digitizer. Data collection begins after the laser 
pulse has terminated to make deconvolution of the laser excitation 
unnecessary. Data are recorded with the laser tuned on and off the 
excitation transition. The decay rate is extracted from the 
background-subtracted fluorescence trace by performing a linear least-square 
fit to a single exponential. This is repeated 5 times to determine the 
lifetime of the level. The lifetime is measured twice for each level, using 
a different excitation transition whenever possible. This redundancy helps 
ensure that the transitions are identified correctly in the experiment, 
classified correctly, and are free from blends.

The lifetimes reported here have an uncertainty of $\pm $5{\%}, except for 
the shortest lifetimes ($<$4 ns) for which the uncertainties are $\pm $0.2 
ns. These uncertainties are primarily systematic, not statistical. The 
possible systematic errors in our measurements must be well understood and 
controlled in order to achieve this level of uncertainty. They include 
limits of the electronic bandwidth, cascade fluorescence, Zeeman quantum 
beats and atomic motion flight-out-of-view effects, among others. The 
dominant systematic error depends on the lifetime. For example the bandwidth 
limits, linearity, and overall fidelity of the electronic detection system 
results in the increasing fractional uncertainty below 4 ns and a lower 
limit of $\sim $2 ns for our system. These systematic effects are discussed 
in detail in earlier publications, (See, for example, Den Hartog et al. 
1999; 2002) and will not be discussed further here. As a means of verifying 
that the measurements are within the stated uncertainties, we perform 
periodic end-to-end tests of the experiment by measuring a set of well-known 
lifetimes. These cross-checks include lifetimes of Be \textsc{i} (Weiss 
1995), Be \textsc{ii} (Yan et al. 1998) and Fe \textsc{ii} (Guo et al. 1992; 
Bi\'{e}mont et al. 1991), covering the range from 1.8--8.8 ns. An Ar 
\textsc{i} lifetime is measured at 27.85 ns (Volz {\&} Schmoranzer 1998). He 
\textsc{i} lifetimes are measured in the range 95 -- 220 ns (Kono {\&} 
Hattori 1984). 

The results of our lifetime measurements of 41 odd-parity levels of Hf 
\textsc{ii} are presented in Table 1. Energy levels are from the tabulation 
by Moore (1971). Air wavelengths are calculated from the energy levels using 
the standard index of air (Edl\'{e}n 1953). The uncertainty of the lifetimes 
is the larger of $\pm $5{\%} or $\pm $0.2 ns.

Also presented in Table 1 is a comparison of our results with those of 
Lundqvist et al (2006), which are the only other LIF lifetime measurements. 
Of the 41 measurements, our results overlap for 17 of the 18 radiative 
lifetimes they report. We see generally good agreement with their results 
for short lifetimes $<$5 ns, but their results are slightly, 5 -- 15{\%}, 
longer than ours for lifetimes in the 10 -- 35 ns range. In the middle of 
the hafnium data taking we remeasured the Ar \textsc{i} cross check which 
Volz and Schmoranzer (1998) measured to be 27.85 ns. We reproduced this 
lifetime within 1{\%}, giving us confidence that we understand the 
systematics in the 20 to 30 ns lifetime range. This small discordance 
between our (UW group) measurements and those by the Lund University group 
of Lundqvist et al. is not serious. The worst disagreement is only 15{\%}, 
which is only slightly beyond our combined uncertainties. Including all 17 
lifetimes for which we overlap, we see a mean difference of +4.5{\%} in the 
sense of ($\tau _{Lund}$ --$\tau_{UW})$/$\tau_{UW} $ 
and a similar root-mean-squared difference of 7.3{\%} 
between our sets of measurements. We further discuss this point in the next 
section.

\begin{center}
3. BRANCHING FRACTIONS AND ATOMIC TRANSITION PROBABILITIES
\end{center}

We primarily used spectra from the 1.0 meter FTS at the National Solar 
Observatory (NSO) for this project on Hf \textsc{ii}. A few supplemental 
measurements on some deep UV lines were recorded in our Univ. of Wisconsin 
lab using a 1.0 m Acton spectrometer system with a photodiode array and a 
small pre-monochromator. The 1.0 meter FTS is our preferred instrument for 
many reasons. It has the large etendue of all interferometric spectrometers, 
a limit of resolution as small as 0.01 cm$^{-1}$, wavenumber accuracy to 1 
part in 10$^{8}$, broad spectral coverage from the UV to IR, and the 
capability of recording a million point spectrum in 10 minutes (Brault 
1976). All instruments of this type are insensitive to any small drift in 
source intensity since an interferogram is a simultaneous measurement of all 
spectral lines.

The energy level structure of Hf \textsc{ii} is relatively simple compared 
to the nearby rare earth elements. The low even-parity levels belong to the 
ground 5d6s$^{2}$ configuration, the 5d$^{2}$6s configuration starting at 
3645 cm$^{-1}$, and the 5d$^{3}$ configuration starting at 18,898 
cm$^{{\-}1}$. Both levels, including the ground a$^{2}$D$_{3/2}$ level, of 
the 5d6s$^{2}$ configuration are known. Only the high-lying $^{2}$S level of 
the 5d$^{2}$6s configuration is unknown, the other 15 levels are known. The 
5d$^{3}$ configuration is also nearly complete, except for the higher of the 
two $^{2}$D terms. The absence of unknown low-lying even-parity levels 
simplified our search for all possible branches from the odd-parity upper 
levels of this study.

The known low-lying odd-parity configurations of Hf \textsc{ii} include the 
5d6s6p configuration starting at 28,069 cm$^{-1}$, and the 5d$^{2}$6p 
configuration starting at 42,518 cm$^{-1}$. A parametric study of Hf 
\textsc{ii} by Wyart {\&} Blaise (1990) did reveal two new levels of these 
odd-parity configurations with energies above 60,000 cm$^{-1}$, but 
described the 6s$^{2}$6p configuration as ``unrecognizable'' due to mixing.

We considered recording new data on Hf \textsc{ii}, but found that existing 
FTS data in the electronic archives of the National Solar Observatory were 
more than adequate for our project.\footnote{ The NSO archives are available 
at http://nsokp.nso.edu/dataarch.html} In order to make our branching 
fraction measurements as complete as possible, we worked on the 13 spectra 
listed in Table 2. A co-author and collaborators recorded some of these Hf 
spectra listed in Table 2 during observing runs in the 1984 period while 
working on Hf \textsc{i }(Duquette et al. 1986). The other spectra from 1987 
listed in Table 2 were recorded by Earl Worden as a guest observer. All 13 
spectra were recorded on custom, water-cooled HCD lamps with either Ar or Ne 
as the buffer gas. A sufficient range of discharge currents was used to 
check the strongest Hf \textsc{ii} lines to low-lying levels for optical 
depth errors. These potential errors were identified and eliminated by 
comparing the high- and low-current HCD spectra. Weaker lines are not 
susceptible to optical depth errors but are more susceptible to error from 
blending with buffer gas lines and Hf \textsc{i }lines. Buffer gas lines, Hf 
\textsc{i }lines, and Hf \textsc{ii }lines all have a different dependence 
on discharge current. Furthermore, the ratio of Hf \textsc{i }line 
intensities to Hf \textsc{ii }line intensities is dependent on buffer gas. 
The comparison of Hf-Ar spectra over a range of currents and the comparison 
of the Hf-Ne spectrum with Hf-Ar spectra were used in eliminating potential 
errors from line blends.

The establishment of an accurate relative radiometric calibration or 
efficiency is critical to a branching fraction experiment. As indicated in 
Table 2, we depended primarily on the Ar \textsc{i} and Ar \textsc{ii} line 
technique. This calibration technique captures the wavelength-dependent 
response of detectors, spectrometer optics, lamp windows, and any other 
components in the light path or any reflections which contribute to the 
detected signal (such as due to light reflecting off the back of the hollow 
cathode). The technique is based on a comparison of well-known branching 
ratios for sets of Ar \textsc{i} and Ar \textsc{ii} lines widely separated 
in wavelength, to the intensities measured for the same lines. Sets of Ar 
\textsc{i} and Ar \textsc{ii} lines have been established for this purpose 
in the range of 4300 to 35000 cm$^{-1}$ (23256 to 2857 {\AA}) by Adams {\&} 
Whaling (1981), Danzmann {\&} Kock (1982), Hashiguchi {\&} Hasikuni (1985), 
and Whaling et al. (1993). 

Tungsten (W) filament standard lamps are particularly useful near the Si 
detector cutoff in the 9,000 to 10,000 cm$^{-1}$ range (11,111 to 10,000 
{\AA}) where the FTS sensitivity is changing rapidly as a function of wave 
number, and near the dip in sensitivity at 12,500 cm$^{-1}$ (8000 {\AA}) 
from the aluminum coated optics. Tungsten lamps are not bright enough to be 
useful for FTS calibrations in the UV region, and UV branches dominate the 
decay of all the Hf \textsc{ii} levels in this study. In general one must be 
careful when using continuum lamps to calibrate the FTS over wide spectral 
ranges, because the ``ghost'' of a continuum is a continuum. The highest 
current spectrum (see {\#}6 of Table 2) that has redundant Ar line and W 
lamp calibrations, was very valuable in measurements of weak visible and 
near IR lines. 

Branching fractions were completed for all odd-parity Hf \textsc{ii} levels 
that do not have deep UV branches beyond the limit of our FTS spectra. This 
list includes all odd-parity levels below 40,000 cm$^{-1}$, except the 
y$^{2}$F$_{5/2}$ at 38578 cm$^{-1}$, and a few odd-parity levels above 
40,000 cm$^{-1}$. Some of the levels, specifically the high z$^{4}$G levels 
and the y$^{2}$D$_{3/2}$ level, have deep UV branches on the FTS spectra but 
beyond the limit of the Ar line calibration. The deep UV branches from these 
levels were measured using the 1.0 m Acton spectrometer system with a 
photodiode array and pre-monochromator as described by Den Hartog et al. 
(2005). The radiometric response of the Acton spectrometer system in the 
deep UV was determined using an Ar mini-arc lamp. Our Ar mini-arc lamp was 
calibrated at NIST and operated without a window (see Bridges {\&} Ott 1977; 
Klose, Bridges, {\&} Ott 1988 for discussions of the Ar mini-arc as a deep 
UV radiometric standard). Ar mini-arcs, when used without a window at 
wavelengths $>$ 2000 {\AA}, have exceptional short- and long-term stability 
as radiometric standards. A small commercially-manufactured, sealed HCD lamp 
was used during measurements with the 1.0 m Acton spectrometer system.

Every possible transition between known energy levels of Hf \textsc{ii} 
satisfying both the parity change and $\Delta $J = -1, 0, or 1 selection 
rules was studied during analysis of FTS data. Energy levels from Moore 
(1971) were used to determine possible transition wave numbers. A subset of 
the Hf \textsc{ii} energy levels have been improved to interferometric 
accuracy by Lundqvist et al. (2006), but the older values from Moore are not 
seriously in error.

We set baselines and integration limits ``interactively'' during analysis of 
the FTS spectra. The same numerical integration routine was used to 
determine the un-calibrated intensities of Hf \textsc{ii} lines and selected 
Ar \textsc{i} and Ar \textsc{ii} lines used to establish a relative 
radiometric calibration of the spectra. A simple numerical integration 
technique was used in this and most of our other recent studies because of 
weakly resolved or unresolved hyperfine and isotopic structure. More 
sophisticated profile fitting is used only when the line sub-component 
structure is either fully resolved in the FTS data or known from independent 
measurements.

The procedure for determining branching fraction uncertainties was described 
in detail by Wickliffe et al. (2000). Branching fractions from a given upper 
level are defined to sum to unity, thus a dominant line from an upper level 
has small branching fraction uncertainty almost by definition. Branching 
fractions for weaker lines near the dominant line(s) tend to have 
uncertainties limited by their S/N ratios. Systematic uncertainties in the 
radiometric calibration are typically the most serious source of uncertainty 
for widely separated lines from a common upper level. We used a formula for 
estimating this systematic uncertainty that was presented and tested 
extensively by Wickliffe et al. (2000). The highest current spectrum of the 
HCD lamp enabled us to connect the stronger visible and near IR branches to 
quite weak branches in the same spectral range. Uncertainties grew to some 
extent from piecing together branching ratios from so many spectra, but such 
effects have been included in the uncertainties on branching fractions of 
the weak visible and near IR lines. In the final analysis, the branching 
fraction uncertainties are primarily systematic. Redundant measurements with 
independent radiometric calibrations help in the assessment of systematic 
uncertainties. Redundant measurements from spectra with different discharge 
conditions also make it easier to spot blended lines and optically thick 
lines. 

Branching fractions from the FTS spectra were combined with the radiative 
lifetime measurements described in {\S}2 to determine absolute transition 
probabilities for 150 lines of Hf \textsc{ii} in Table 3. Air wavelengths in 
Table 3 were computed from energy levels (Moore 1971) using the standard 
index of air (Edl\'{e}n 1953). 

Transition probabilities for the very weakest lines that were observed with 
poor S/N ratios (branching fractions $<$ 0.001), and for a few blended 
lines, are not included in Table 3; however these lines are included in the 
branching fraction normalization. The effect of the problem lines becomes 
apparent if one sums all transition probabilities in Table 3 from a chosen 
upper level, and compares the sum to the inverse of the upper level lifetime 
from Table 1. Typically the sum of the Table 3 transition probabilities is 
between 98{\%} and 100 {\%} of the inverse lifetime. Although there is 
significant fractional uncertainty in the branching fractions for these 
problem lines, this does not have much effect on the uncertainty of the 
stronger lines that were kept in Table 3. Branching fraction uncertainties 
are combined in quadrature with lifetime uncertainties to determine the 
transition probability uncertainties in Table 3. Possible systematic errors 
from missing branches to unknown lower levels are negligible in Table 3, 
because we were able to make at least rough measurements on visible and near 
IR lines with branching fractions as small as 0.001 . The generally short Hf 
\textsc{ii} lifetimes, in combination with the frequency cubed scaling of 
transition probabilities, means that any unknown line in the mid- to far-IR 
region will not have a significant branching fraction. 

The recently published work by Lundqvist et al. (2006) provides a valuable 
opportunity for comparing at least some of our branching fraction 
measurements to independent, modern branching fraction measurements. 
Branching fraction uncertainties are primarily systematic, not statistical. 
It has, in most comparisons, not been possible for independent research 
groups to achieve the level of agreement in branching fraction measurements 
that is routinely achieved in LIF radiative lifetime measurements (e.g. 
Lawler et al. 2006). Both our branching fraction experiment and the 
experiment by Lundqvist et al. employed high performance FTS's. The Chelsea 
Instruments FT 500 at Lund University has better deep UV performance than 
the 1.0 m FTS at Kitt Peak, but there is enough overlap between our UW 
experiment and the Lund experiment for a meaningful comparison. The two 
experiments utilized different spectrometers, different lamps, and different 
analysis software. There is, however, some overlap in calibration technique. 
Lundqvist et al. used the Ar line technique for wavenumbers below 30,000 
cm$^{-1}$ (wavelengths above 3333 {\AA}) and a D$_{2}$ standard lamp for 
wavenumbers above 27,800 cm$^{-1}$ (wavelengths below 3597 {\AA}). This 
comparison cannot be considered to be a test of the Ar \textsc{i} and Ar 
\textsc{ii} calibration lines, but as mentioned above there have been 
multiple independent test of the Ar branching ratios. Figure 1 is a plot of 
the branching fractions from the Lund University effort divided by our 
University of Wisconsin (UW) branching fraction as a function of wavelength 
for 72 transitions from eight upper levels. Figure 2 is a similar plot as a 
function of the UW branching fraction. The error bars in this plot were 
determined by combining in quadrature the UW relative branching fraction 
uncertainty with the relative Lund transition probability uncertainty. 
Lundqvist et al. did not report separate branching fraction uncertainties. 
For a weaker line the branching fraction uncertainty typically dominates the 
transition probability uncertainty, but for strong lines the lifetime 
uncertainty dominates the transition probability uncertainty. This means 
that the error bars are somewhat larger than desired for the strong UV 
transitions. The comparison reveals good (single error bar) agreement on all 
but a few branching fractions. If the data Figure 1 is replotted with points 
of similar wavelength averaged together, then a weak dependence on 
wavelength becomes visible. This wavelength dependence is probably due to 
slightly (10{\%} to 15{\%}) different radiometric calibrations over the 3000 
to 6500 ? range. Such differences over more than a factor of 2 in wavelength 
are consistent with estimated uncertainties in the calibration techniques. A 
close inspection of Figure 2 reveals greater discordance for the weaker 
branches. This is expected in part because uncertainty migrates to the 
weaker branches by the definition of a branching fraction as discussed 
above. The weaker branches are also more vulnerable to line blending errors 
and S/N limitations.

Figure 3 is a comparison of log(\textit{gf}) values from the Lund University study to 
our (UW) log(\textit{gf}) values with Delta log(\textit{gf}) = log(\textit{gf})$_{Lund}$ -- log(\textit{gf})$_{UW}$ 
plotted as a function of log(\textit{gf})$_{UW}$. The effect of the slight differences 
in radiative lifetime measurements on the longer lived levels as discussed 
in {\S} 2 is visible; the average Delta log(\textit{gf}) is clearly less than zero. 
Never-the-less most of the Delta log(\textit{gf}) are within one error bar of zero. 
Lundqvist et al. (2006) compared their measurements to log(\textit{gf}) values from the 
beam foil study by Andersen et al. (1976) and to log(\textit{gf}) values from the arc 
emission study by Corliss {\&} Bozman (1962). We omit those comparisons 
here, since the above comparison to Lundqvist et al.'s results from laser 
and FTS measurements is the most relevant. 

\begin{center}
4. SOLAR AND STELLAR HAFNIUM ABUNDANCE
\end{center}

We have employed the new Hf \textsc{ii} transition probabilities to 
re-determine the hafnium abundance of the solar photosphere, and to derive 
its abundance in 10 very metal-poor ([Fe/H] $\le $ -1.5)\footnote{ We adopt 
standard stellar spectroscopic notations that for elements A and B, [A/B] = 
log$_{10}$(N$_{A}$/N$_{B})_{star}$ - log$_{10}$(N$_{A}$/N$_{B})_{sun}$, 
for abundances relative to solar, and log $\varepsilon $(A) = 
log$_{10}$(N$_{A}$/N$_{H})$ + 12.0, for absolute abundances.} stars that 
have large overabundances of the rare-earth elements (e.g., [Eu/Fe] $\ge $ 
+0.5). Our analyses followed the methods used in previous papers of this 
series (in particular Lawler et al. 2006 and Den Hartog et al. 2006, 
hereafter DLSC06).

\begin{center}
4.1 Line Selection
\end{center}

We have determined accurate transition probabilities for 150 Hf \textsc{ii} 
lines, but unfortunately very few of these can be used in abundance analyses 
of the Sun and cool metal-poor stars. This can be understood by considering 
line strength factors for the Hf \textsc{ii} transitions. In a standard LTE 
abundance analysis the relative strengths of lines of individual species 
vary directly as the product of elemental abundances, transition 
probabilities, Saha ionization corrections, and Boltzmann excitation 
factors. For Hf \textsc{ii}, like all of the nearby rare-earth first ions, 
the problem is simplified, because the ionization potential is relatively 
low (6.825 eV for Hf, Grigoriev {\&} Melikhov 1997). In photospheres of the 
Sun and red giant stars all these elements are completely ionized, or n$_{II 
}\approx $ n$_{total}$. Therefore Saha corrections to account for other 
ionization state populations are very small and can be ignored. Then for a 
weak line on the linear part of the curve-of-growth, the equivalent width 
(EW) and reduced width (RW) are related as log(RW) $\equiv $ 
log(EW/$\lambda $)  $\propto$  log($\epsilon$\textit{gf}) -- $\theta$$\chi$, 
where $\epsilon $ is the elemental abundance, \textit{gf} is the oscillator 
strength, $\chi $ is the excitation energy in units of eV, and 
$\theta \equiv $ 5040/T is the inverse temperature. We thus define the 
relative strength factor of a transition, ignoring line saturation effects, 
as STR $\equiv $ log($\epsilon $\textit{gf}) -- $\theta \chi $. 
Ionized-species transitions of elements with low first ionization potentials 
can be reliably inter-compared with these strength factors.

In Figure 4 we plot Gd \textsc{ii} and Hf \textsc{ii} line strength factors 
as a function of wavelength. The left-hand panel, showing Gd \textsc{ii} 
data, is identical to the right-hand panel of Figure 2 in DLSC06. The 
right-hand panel of Figure 4 is generated from the Hf \textsc{ii} data of 
this paper. For these computations we have assumed $\theta $ = 1.0, a 
compromise value between that of the Sun (T$_{eff}$ = 5780 K or $\theta $ = 
0.87) and metal-poor giants (T$_{eff}  \approx $ 4600 K or $\theta $ 
$\approx $ 1.10). The exact value of $\theta $ is not important for our 
purposes, since the vast majority of measurable Gd \textsc{ii} and Hf 
\textsc{ii} lines in the Sun and metal-poor stars of interest here arise 
from low excitation energy states, $\chi   <$ 1 eV). We have adopted 
approximate solar abundances of log $\varepsilon $(Gd) = +1.1 (DLSC06), and 
log $\varepsilon $(Hf) = +0.9 (close to the photospheric value suggested by, 
e.g,, Lodders 2003, which we will confirm in this study) . 

In Figure 4 we have also indicated the approximate strength factors for 
extremely weak Gd \textsc{ii} and Hf \textsc{ii} lines in the solar 
spectrum, and for lines that are reasonably strong. The assignment of these 
strength levels is discussed in detail by DLSC06. Briefly, the very weak 
line strength level was estimated by first assuming that the weakest 
unblended lines routinely measurable on the Delbouille et al. (1973) 
center-of-disk solar spectrum have EW$_{weak} \approx $ 1.5 m{\AA} in the 
blue spectral region ($\lambda \sim $ 4500 {\AA}), or log(RW)$_{weak}$ 
$\approx $ -6.5. Repeated searches for weak lines of Sm \textsc{ii} (Lawler 
et al. 2006) and Gd \textsc{ii} (DLSC06) indicated that log(RW)$_{weak}$ 
$\approx $ -6.5 corresponds to STR$_{weak} \approx $ -0.6. This 
very-weak-line strength estimate will also apply to Hf \textsc{ii} 
transitions. The strong-line strength value was more arbitrarily set at a 
value 20 times larger, or STR$_{strong}$ = -0.6 + 1.3 = +0.7. If lines 
remained unsaturated then their equivalent widths would scale linearly: 
log(RW)$_{strong} \approx $ -5.2, or EW$_{strong} \approx $ 30 m{\AA} 
near 4500 {\AA}. Lines of this strength are somewhat saturated, such that in 
the solar spectrum STR $\approx $0.7 corresponds to log(RW) 
$\approx $ -5.35, or EW $\approx $ 20 m{\AA} at 4500 {\AA}. 

To summarize the above discussion, lines of Sm \textsc{ii}, Gd \textsc{ii}, 
and Hf \textsc{ii} (as well as lines of other ionized rare-earth elements) 
with STR $\approx $ -0.6 are so weak that they are difficult to detect in 
the solar spectrum, and those with STR $>$ +0.7 are strong. Using these 
line-strength criteria, Gd \textsc{ii} has about 250 potential lines for 
solar abundance analyses, and nearly 50 strong lines. In contrast, as is 
obvious from Figure 4, Hf \textsc{ii} simply has very few promising 
transitions: only 21 lines with STR $>$ -0.6, and no strong lines at all. 
These numbers are qualitatively consistent with solar spectrum 
identifications given by Moore, Minnaert, {\&} Houtgast (1966): 60 lines of 
Gd \textsc{ii} but only 18 of Hf \textsc{ii}.

We repeated the procedures described in DLSC06 and earlier papers of this 
series to identify the final set of Hf \textsc{ii} lines to be used in the 
solar/stellar abundance analyses. Having relatively few potential Hf 
\textsc{ii} lines, we carefully considered all with STR $>$ -1.5. From 
visual comparison of the electronic version\footnote{Available at 
http://bass2000.obspm.fr/solar{\_}spect.php} of the Delbouille et al. (1973) 
solar center-of-disk spectrum with the spectrum of the $r-$process-rich 
metal-poor giant star \bd17 (Cowan et al. 2002), and from review of the 
Moore et al. (1966) solar line identifications, we eliminated the completely 
unsuitable lines: those that are undetectably weak and/or severely blended 
in both of these spectra. We then consulted the Kurucz (1998) atomic and 
molecular line compendium, in order to eliminate remaining candidate lines 
that suffer significant contamination by transitions of other 
neutron-capture species. In the end only 19 Hf \textsc{ii} lines survived to 
be subject to closer inspection in solar and/or stellar spectra. 

\begin{center}
4.2 The Solar Photospheric Hafnium Abundance
\end{center}

We employed synthetic spectrum computations following procedures discussed 
by DLSC06. Atomic and molecular line lists in 4 to 6 {\AA} intervals 
surrounding each Hf \textsc{ii} transition were compiled from Kurucz's 
(1998) line database and Moore et al.'s (1966) solar identifications. These 
line lists and the Holweger {\&} M\"{u}ller (1974) solar model atmosphere 
were used as inputs into the current version of the LTE line analysis code 
MOOG (Sneden 1973) to generate the synthetic spectra. The Solar atmospheric 
model parameters are listed in Table 4. We assumed the solar photospheric 
abundances recommended in reviews by e.g, Grevesse {\&} Sauval (1998, 2002) 
and Lodders (2003), as well as values for neutron-capture elements 
determined in earlier papers of this series. Transition probabilities for 
ionized species of neutron-capture elements were taken from these recent 
studies: Y, Hannaford et al. (1982); Zr, Malcheva et al. (2006); La, Lawler 
et al. (2001a); Ce, Palmeri et al. (2000); Pr, Ivarsson et al. (2001); Nd, 
Den Hartog et al. (2003); Sm, Lawler et al. (2006); Eu, Lawler et al. 
(2001b); Gd, DLSC06; Tb, Lawler et al. (2001c); Dy, Wickliffe et al. (2000); 
Ho, Lawler et al. (2004); and Hf, the present work. 

We computed multiple synthetic spectra for the 19 selected Hf lines, and 
compared them to the Delbouille et al. (1973) photospheric spectrum. The 
synthetic spectra were smoothed with a Gaussian to match the observed line 
broadening (from photospheric macroturbulence and instrumental effects). As 
in previous papers in this series, predicted contaminant absorptions were 
matched to the solar spectrum by: (a) altering oscillator strengths for 
known atomic transitions, except for the species listed above; (b) varying 
the abundances of C, N, and/or O for CH, CN, NH, and OH band lines: and (c) 
adding Fe \textsc{I} lines with excitation potentials $\chi $ = 3.5 eV and 
arbitrary transition probabilities for unknown absorptions. After 
modifications of the line lists to match the solar spectrum, similar trials 
were also performed for \bd17. These initial synthetic spectrum 
computations demonstrated that 7 of the 19 proposed Hf \textsc{ii} lines 
were useless for abundance determinations in the Sun and even in the most 
$r-$process-rich of the metal-poor stars discovered to date. The final set of 12 
Hf \textsc{ii} lines remaining after these tests is listed in Table 5.

In the solar photospheric spectrum we found just four Hf \textsc{ii} 
features to be good hafnium abundance indicators. These are displayed in 
Figure 5. Inspection of this figure illustrates the analytical problems 
discussed above: the Hf \textsc{ii} lines are all weak and blended to 
varying degrees. In fact, only the 4093.15 {\AA} line (panel a) is largely 
unblended and occurs in a wavelength region interval where the solar 
continuum can be defined reasonably well. The 3918.09 {\AA} line (panel b) 
has a weak local blend by Co \textsc{i} 3918.06 {\AA}, but more importantly 
the continuum level is defined by an Fe \textsc{i} blend near 3918.5 {\AA} 
and an extremely strong Fe \textsc{i} line at 3920.26 {\AA}. The 3561.66 
{\AA} line (panel c) should be one of the strongest Hf \textsc{ii} 
transitions; see Figure 4. In reality it is a minor constituent of the blend 
dominated by Ti \textsc{ii} 3561.58 {\AA} and Ni \textsc{i} 3561.76 {\AA}. 
Finally, the 3389.83 {\AA} line (panel d) is clearly present, but its 
intrinsic weakness and proximity to the strong Fe \textsc{i} 3389.75 {\AA} 
line renders it useful only in confirming the hafnium abundance determined 
from the other features.

Abundances derived for these Hf \textsc{ii} lines are listed in Table 5. A 
straight mean abundance is log $\varepsilon $(Hf) = 0.88 $\pm $ 0.02 
($\sigma $ = 0.03), where $\sigma $ is one standard deviation of the 
set of measurements. The formal scatter figure surely must underestimate the 
actual internal errors, because the synthetic/observed spectrum fitting 
procedures needed several judgment decisions to account for significant line 
blending issues for all but the 4093 {\AA} line. Repeated test syntheses of 
the four lines, altering various feature contamination and continuum 
placement assumptions, suggest that a more conservative estimate of log 
$\varepsilon $(Hf) = 0.88 $\pm $ 0.08 is appropriate if all four lines are 
given equal weight. If the hafnium abundance were to be based only on the 
4093 {\AA} line then we estimate log $\varepsilon $(Hf) = 0.86 $\pm $ 0.05. 

Since Hf \textsc{ii} lines arise from ionization/excitation conditions that 
are very similar to those of nearby rare-earth species studied in this 
series of papers, all the same external (scale) errors discussed by DLSC06 
for Gd \textsc{ii} apply here. The hafnium abundance varies directly with 
the Hf \textsc{ii} partition, but the polynomial representation of Irwin 
(1981) agrees well with our calculations from currently-available energy 
level data as discussed in {\S} 3. Within the analysis assumptions of LTE 
and plane-parallel model atmospheric geometry, the choice of solar model 
photosphere (e.g., Kurucz 1998 or Grevesse {\&} Sauval 1999 instead of the 
Holweger {\&} M\"{u}ller 1974 model used here) changes the derived hafnium 
abundance by only 0.01 to 0.02 dex. Application of more rigorous modeling of 
the atomic and solar atmospheric physics should be done in the future, for 
hafnium and the nearby rare-earth elements.

Andersen et al. (1976) performed the most comprehensive previous solar 
hafnium abundance study. They derived log $\varepsilon $(Hf) = 0.88 $\pm $ 
0.08 from eight Hf \textsc{ii} transitions, obviously in excellent agreement 
with our result. They used their own Hf \textsc{ii} lifetime data to correct 
the transition probabilities of Corliss {\&} Bozman (1962), and employed 
synthetic spectra in their solar abundance analysis. Our lab analysis 
includes all of their lines, for which our transition probabilities are 
systematically lower: $\Delta $log(\textit{gf}) = -0.10, in the sense this study minus 
Andersen et al. A simple correction to their abundance would shift it to log 
$\varepsilon $(Hf) = 0.98. However, our solar analysis suggests that most of 
the Andersen et al. lines are too blended to yield good solar hafnium 
abundances. Only the 3389 and 3561 {\AA} lines are in common with the 
present work, and these are our two less reliable lines. Using just these 
lines, updating the Andersen et al. \textit{gf'}s to our values, would yield log 
$\varepsilon $(Hf) = 0.96, not in serious disagreement with our work. 

The Lundqvist et al. (2006) laboratory Hf \textsc{ii} study did not perform 
a detailed solar abundance analysis. Their transition probability scale is 
smaller than that of Andersen et al. (1976): $\Delta $log(\textit{gf}) $\approx$ -0.15. This led them to suggest that the Andersen et al. solar 
hafnium abundance, ``... may be underestimated by 0.1 and 0.2 dex'', i.e. 
log $\epsilon $(Hf) $\approx$ 1.0. Direct application of the Lundqvist 
et al. gf's to our solar abundance analysis is not possible because none of 
our four photospheric transitions were included in their work. If the 
generally small mean offset between their transition probability scale and 
ours were applied to these lines, our photospheric abundance would become 
log $\varepsilon $(Hf) = 0.88 + 0.04 = 0.92, in agreement with our 
recommended value within the uncertainties of our study and theirs. 

The photospheric hafnium abundance is somewhat larger than the current best 
meteoritic estimates, e.g, log $\varepsilon $(Hf) = 0.77 $\pm $ 0.04 
(Lodders 2003), but the error bars of the two values do overlap. We will 
consider the consequences of this small discrepancy in {\S}5. 

\begin{center}
4.3 Hafnium Abundances in Metal-Poor, $r-$Process-Rich Stars
\end{center}

We have determined hafnium abundances in 10 very metal-poor giant stars that 
are known to be enriched in products of $r-$process nucleosynthesis. In past 
papers of this series we have only analyzed three extreme cases, CS 
22892-052, \bd17, and HD 115444. However, hafnium is an important 
element in connecting the lightest rare-earth elements (e.g. La, Z = 57) 
with elements of the 3$^{rd} r-$process peak (Os, Ir, Pt, Z = 76 to 78). To 
our knowledge, hafnium is the heaviest (Z = 72) stable element represented 
by low-excitation ($\chi<$ 1.5 eV) ionized lines in cool, metal-poor 
stellar spectra. The rare earths and the long-lived chronometer elements Th 
and U also are detected only via these types of transitions. Derived 
abundance ratios Th/(rare earth) and Hf/(Th or U) are largely insensitive to 
uncertainties in model atmosphere parameters including: effective 
temperature, gravity, microturbulent velocity, and overall metallicity. 
Therefore we have expanded the star list to include 10 $r$-process-rich stars 
that have been studied in other abundance surveys (Burris et al. 2000, 
Simmerer et al. 2004, Cowan et al. 2005). 

We determined Hf abundances from as many of the 12 candidate lines as 
possible for each of the program stars in the same manner as was done for 
the Sun. In Table 4 we list the model atmosphere parameters and their 
sources. The individual Hf \textsc{ii} line abundances are listed in Table 
5. The mean Hf abundances, uncertainties, and number of lines used are 
listed in Table 6.

We also list mean abundances for La and Eu in Table 6. For HD 221170, \bd17, 
 and CS 22892-052 these values were adopted from the original papers. 
For the remaining stars we chose to re-derive La and Eu abundances from 
synthetic spectrum analyses of up to nine La \textsc{ii} and six Eu 
\textsc{ii} lines in the spectral region 3700 {\AA} $<  \lambda<$ 4450 
{\AA}. Atomic data for the La and Eu transitions were taken from Lawler et 
al. (2001a) and Lawler et al. (2001b), respectively. The new abundances are 
in good agreement with previously published values. There are six stars in 
common with Simmerer et al. (2004). Defining differences in the sense this 
study \textit{minus} Simmerer et al., we find $\Delta $ log $\varepsilon $(La) = -0.06 
$\pm $ 0.02 ($\sigma $ = 0.06), $\Delta $ log $\varepsilon $(Eu) = -0.01 
$\pm $ 0.02 ($\sigma $ = 0.04), and $\Delta $ log $\varepsilon $(La/Eu) = 
-0.04 $\pm $ 0.01 ($\sigma $ = 0.03). For CS 31082-001, $\Delta $ log 
$\varepsilon $(La/Eu) = -0.06 with respect to the comprehensive Hill et al. 
(2002) work. Finally, for HD 115444 we find $\Delta $ log $\varepsilon 
$(La/Eu) = -0.03 with respect to Westin et al. (2000); note that their work 
on this star predates publication of the improved transition probability 
data for La and Eu.

Few Hf abundances have been derived for very metal-poor stars. Lundqvist et 
al. (2006) comment on the application of their Hf \textsc{ii} \textit{gf}-values to 
previous work on two very metal-poor $r$-process-rich stars, but do not perform 
independent new analyses. Adjusting the Hill et al. (2002) and Sneden et al. 
(2003) abundances for CS 31082-001 and CS 22892-052, they recommend log 
$\varepsilon $(Hf) = -0.90 and -0.75 respectively. Our new values of -0.88 
and -0.72 (Table 6), are clearly in good agreement with the Lundqvist et al. 
recommendations.

\begin{center}
5. $r-$PROCESS NUCLEOSYNTHESIS AND HAFNIUM ABUNDANCES
\end{center}

Abundances of neutron-capture elements in low metallicity stars, formed 
early in the history of the Galaxy, are predominantly the result of 
$r-$process nucleosynthesis. The site for this synthesis process is presumably 
supernova explosions of high-mass (short-lived) stars. In contrast, slow (or 
$s-$process) nucleosynthesis occurs in low-mass (and long-lived) stars (see 
Cowan {\&} Sneden 2006 for discussion). The $r-$process rich ejecta from early 
supernovae were thus injected into the interstellar medium relatively 
shortly before the formation of the observed (very low metallicity) halo 
stars. It took much longer for the $s-$process material to be incorporated into 
gas that formed somewhat younger and higher metallicity stars (Burris et al. 
2000; Simmerer et al. 2004). 

The relative neutron-capture abundance distributions exhibit clear 
star-to-star consistency in very metal-poor halo stars. This has suggested, 
for example, a robust $r-$process operating over billions of years and a unique 
site, or at least a unique set of synthesis conditions for these elements 
(Cowan {\&} Sneden 2006). Abundance determinations, particularly from high 
S/N spectra coupled with very precise atomic data, and comparisons among the 
halo stars can therefore provide critical new information about the 
synthesis mechanisms and sites for the $r-$process. Furthermore, since the halo 
distributions appear to be scaled solar $r-$process, these abundance comparisons 
can be employed to constrain the $r-$process and $s-$process-only solar system 
distributions. 

\begin{center}
5.1 Solar and Stellar Abundance Comparisons
\end{center}

In the 10 $r-$process rich stars in our sample, hafnium exhibits near-constant 
abundances with respect to La and Eu: log $\varepsilon $(Hf/La) = -0.13 $\pm 
$ 0.02 ($\sigma $ = 0.06) and log $\varepsilon $(Hf/Eu) = +0.04 $\pm $ 0.02 
($\sigma $ = 0.06). We show the variation of log $\varepsilon $(La/Eu) in 
Figure 6 and log $\varepsilon $(Hf/Eu) in Figure 7 as a function of [Fe/H] 
and [Eu/H]. The star-to-star scatter for both log $\varepsilon $(La/Eu) and 
log $\varepsilon $(Hf/Eu) is consistent with observational/analytical 
uncertainties. No trends with metallicity or overall $r$-process enhancement 
are obvious, probably partly as a consequence of the selection of the 
(mostly all very low-metallicity) stars to include here.

We first consider log $\varepsilon $(La/Eu) as a function of metallicity in 
Figure 6. We have chosen our target stars to be $r-$process rich. One such 
indication of this is the value of log $\varepsilon $(La/Eu). While La is 
predominantly an $s-$process element in solar material, at early Galactic times 
prior to the onset of the bulk of $s-$process production, La was synthesized in 
the $r-$process. Typical values for $r-$process rich stars are found to be log 
$\varepsilon $(La/Eu) $\approx $ 0.1, while in solar system material (with a 
large $s-$process contribution to La) this value is typically about 0.7 
(Simmerer et al. 2004, Cowan et al. 2006). It is seen in Figure 6 that there 
is a consistency of values for the 10 target stars, with an average of 0.17 
(indicated by the solid line) - a clear indication of the $r-$process rich 
nature of these stars. We note that some of our stars have metallicities 
larger than -2. Some evidence has been found for the $s-$process at 
metallicities as low as -2.5, although the bulk of Galactic \textit{s{\-}}process 
nucleosynthesis appears to occur closer to [Fe/H] $\ge $ -2 (Burris et al. 
2000, Simmerer et al. 2004). We have examined log $\varepsilon $(La/Eu) for 
5 of the most metal-poor ([Fe/H] $<$ -2.5) stars and find almost no 
difference in average value with respect to that found for all 10 of the 
target stars. For comparison we also show in this figure a range of previous 
predictions for the $r-$process-only log $\varepsilon $(La/Eu). The lowest 
dotted line in the figure is a determination based upon earlier 
deconvolutions of the Solar System abundances into $r-$ and $s-$abundances (Simmerer 
et al. 2004, Cowan et al. 2006). These elemental separations are obtained by 
first determining individual isotopic $s-$process contributions in the so called 
classical model approximation (K\"{a}ppeler et al. 1989) or a more 
complicated stellar model approach (Arlandini et al. 1999). (Nuclear data, 
such as neutron-capture cross sections, can in general be obtained for the 
$s-$process nuclei as they are close to stability and have relatively long 
half-lives. In contrast $r-$process nuclei are so radioactive and have such 
short half-lives, that in general, their nuclear properties cannot now be 
experimentally determined.) The $r-$process isotopic abundances are then 
obtained by subtracting the calculated $s-$process isotopic abundances from the 
total solar system abundances, and the elemental $r-$process-only distributions 
are just the sums of the isotopic contributions. Thus, the $r-$process 
abundances are actually residuals and depend very sensitively on the 
$s-$process determinations. Also included in the range of values for the 
predicted $r-$process-only log $\varepsilon $(La/Eu) are those determined based 
upon recent neutron cross section measurements of $^{139}$La from O'Brien et 
al. (2003) and Winckler et al. (2006) -- the latter denoted by the topmost 
dotted line in Figure 6. 

In Figure 7 we show a similar plot of log $\varepsilon $(Hf/Eu) for our 
target stars. In this case we find a mean value of log $\varepsilon $(Hf/Eu) 
= 0.04. Employing only the most metal-poor stars, again makes almost no 
difference in this average. The dotted line indicates the $r-$process predicted 
ratio from Simmerer et al. (2004) and Cowan et al. (2006). 

Despite the consistency of the abundance data in the halo stars with the 
scaled solar system $r-$only distribution, it is seen in both Figures 6 and 7 
that these previous determinations of the (published) solar system $r-$only 
values fall below the log $\varepsilon $(el/Eu) abundance ratios for our 10 
sample stars. This suggests a reexamination of the $r-$only-values for log 
$\varepsilon $(Hf/Eu) and log $\varepsilon $(La/Eu) is required employing 
the more accurate halo abundance determinations, based upon the newly 
measured and more precise atomic lab data. These abundance data can be 
utilized to determine directly the solar system $r-$only values, as opposed to 
obtaining residuals as described above. Analogously to what was done in 
DLSC06, we have first averaged the log $\varepsilon $(el/Eu) offsets (log 
$\varepsilon $(La/Eu) in Figure 6 and log $\varepsilon $(Hf/Eu) in Figure 
7). Europium was used for comparison as it is formed almost entirely in the 
$r-$process, in contrast to La and Hf which have significant contributions from 
the $s-$process (Simmerer et al. 2004). In this manner, we determined the 
expected solar system $r-$only values for both La and Hf. The mean values of the 
ratios for log $\varepsilon $(La/Eu) and log $\varepsilon $(Hf/Eu) in the 10 
stars are shown in the figures with a solid line. Previous determinations of 
the individual $r-$process and $s-$process contributions for La (N$_{r}$ = 0.11 and 
N$_{s}$ = 0.337) and Hf (N$_{r}$ = 0.081 and N$_{s}$ = 0.076) (based upon 
the N$_{Si }$= 10$^{6}$ scale) have been listed by Cowan et al. (2006). We 
find, utilizing the halo star abundances, that the values of N$_{r}$ should 
be revised slightly upward such that N$_{r}$ = 0.134 for La and N$_{r}$ = 
0.099 for Hf. Summing these $r-$process predicted contributions with the 
previously determined $s-$process values yields total solar abundances of 
N$_{tot}$ = 0.471 (log $\varepsilon $ = 1.21) and 0.175 (log $\varepsilon $ 
= 0.782) for La and Hf respectively, again based on the N$_{Si }$= 10$^{6}$ 
scale. The predicted solar system total values for log $\varepsilon $(La/Eu) 
and log $\varepsilon $(Hf/Eu) are shown in Figures 6 and 7 as dashed lines. 
We estimate that the observational and analytic errors probably limit our 
results to the order of $\pm $ 10{\%}. We note that the value we find for 
the total Hf abundance using this technique, log $\varepsilon $(Hf) = 0.782, 
is almost identical to that listed by Lodders (2003) for the meteoritic and 
recommended solar abundance for this element, log $\varepsilon $(Hf) = 0.77 
$\pm $ 0.04. Both values are somewhat below the published photospheric and 
our new determination, log $\varepsilon $(Hf) = 0.88 $\pm $ 0.08, using a 
purely spectroscopic approach. In the case of log $\varepsilon $(La/Eu) our 
predicted total = 0.71 compares well with the Lodders (2003) value (0.66) 
and our observed average solar ratio reported in this paper of 0.62.

In attempting to employ the halo stars for determining the $r-$process-only 
elemental values we are assuming that there are no $s-$process contributions. 
Previous studies have demonstrated remarkable consistencies between the 
detailed stellar abundance distributions (for Ba and above) with the solar 
system $r-$only distribution -- there appear to be no $s-$process contributions and 
none would be expected from the main $s-$process coming from long-lived, 
low-mass stars (see Cowan and Sneden 2006 and references therein). While the 
weak $s-$process (resulting from the capture of neutrons on iron seed nuclei 
during core helium burning in massive stars) can contribute to the lighter 
elemental abundances (up to approximately Zr) in solar metallicity stars, 
this mechanism is not effective in very low-metallicity stars (Pignatari et 
al. 2006). 

\begin{center}
5.2 Hafnium and Nucleocosmochronometry Studies
\end{center}

Long-lived radioactive elements such as Th and U can be employed to 
determine the ages of the oldest stars. Ideally, having the ratio of Th/U 
would provide the most reliable chronometer. However, despite the detection 
of U in CS 31082-001 (Cayrel et al. 2001) and the probable detection of U in 
\bd17 (Cowan et al. 2002), U is difficult to detect -- it is inherently 
weak and the atomic line is often blended with broader molecular lines. 
Th/Eu has been employed for a number of radioactive-age studies, but the 
wide separation in mass number (151 to 153 versus 232) and the (possible) 
associated differences in nucleosynthesis origin has been considered a 
problem for sometime (see recent discussion in Ivans et al. 2006 and 
references therein). In at least one case (i.e., CS 31082-001) Th/Eu also 
gives an unrealistic age estimate (Hill et al. 2002, Schatz et al. 2002). 
The heaviest stable elements, nearby to the actinides, are the third 
$r-$process peak elements Os, Ir, and Pt and the heavier elements Pb and Bi. 
Together, these could be employed in chronometer pairs. Unfortunately, these 
elements are difficult to observe from the ground and Bi has not been 
detected in any metal-poor ($r-$process rich) field halo star.

Recent theoretical work (Kratz et al. 2006), designed to reproduce the total 
solar system $r-$process abundance distribution, has found that the predicted 
abundances of inter-peak element Hf (Z = 72, A = 176 to 180) follow closely 
those of third-peak elements (Os through Pt) and Pb. Thus, Hf, observable 
from the ground, offers promise as a heavy stable element in a chronometer 
pair such as Th/Hf. Therefore, the newly determined Hf abundances in these 
10 halo stars, many of which also have detectable Th, could be employed to 
determine more accurate stellar ages.

\begin{center}
6. CONCLUSIONS
\end{center}
Hafnium (Z = 72) is well suited to be used as a stable reference element in 
nucleocosmochronometry based on the unstable elements Th (Z = 90) and U (Z = 
92) whose radioactive decay timescales are well determined. Hafnium is 
heavier than alternate reference elements such as Nd (Z = 60) and Eu (Z = 
63) and is significantly closer to the 3$^{rd} r-$process peak (Os, Ir, Pt, Z 
= 76 to 78). Multiple lines of Hf \textsc{ii} are measurable in metal poor 
halo stars with enhanced $r-$process abundances. In order to improve Hf 
abundance determinations, we have performed radiative lifetime measurements 
using time-resolved laser-induced fluorescence on 41 odd-parity levels of Hf 
\textsc{ii}. These results were combined with branching fractions from 
Fourier transform spectra to determine absolute atomic transition 
probabilities for 150 lines of Hf \textsc{ii}. Approximately half of our 
measurements overlap with recent independent transition probability 
measurements using the LIF plus FTS method and good agreement is found in a 
detailed comparison. These new laboratory data were applied to (re)determine 
the Hf abundances of the Sun and 10 metal poor, $r-$process rich halo stars. Our 
refined Solar Hf abundance determination, log $\varepsilon $(Hf) = 0.88 $\pm 
$ 0.08, is in agreement with earlier work. The $r$-process-rich stars possess 
constant La/Eu and Hf/Eu abundance ratios to within the uncertainties of our 
analysis; log $\varepsilon $(Hf/La) = -0.13 $\pm $ 0.02 ($\sigma $ = 0.06) 
and log $\varepsilon $(Hf/Eu) = +0.04 $\pm $ 0.02 ($\sigma $ = 0.06). The 
newly determined Hf values set the stage for improved radioactive stellar 
age determinations using the Th/Hf chronometer pair. The observed average 
stellar abundance ratio of Hf/Eu and La/Eu is larger than previous estimates 
of the solar system $r-$process-only value, suggesting a somewhat larger 
contribution from the $r-$process to the production of Hf and La.

\begin{center}
ACKNOWLEDGMENTS
\end{center}
We thank K. Lodders for helpful discussion. This work is supported by the 
National Science Foundation under grants AST- 0506324 (JEL and EADH), 
AST-0307495 (CS), and AST-0307279 (JJC). J. E. Lawler is a guest observer at 
the National Solar Observatory and he is indebted to Mike Dulick and Detrick 
Branstron for help with the 1 m Fourier transform spectrometer. 

\begin{center}
REFERENCES
\end{center}

Adams, D. L., {\&} Whaling, W. 1981, J. Opt. Soc. Am., 71, 1036

Andersen, T., Petersen, P., {\&} Hauge, O., 1976, Sol. Phys., 49, 211

Arlandini, C., K\"{a}ppeler, F., Wisshak, K., Gallino, R., Lugaro, M., 
Busso, M., {\&} Straniero, O. 1999, ApJ, 525, 886

Bi\'{e}mont, E., Baudoux, M., Kurucz, R. L., Ansbacher, W., {\&} Pinnington, 
E. H. 1991,
A{\&}A, 249, 539

Brault, J. W. 1976, J. Opt. Soc. Am., 66, 1081

Bridges, J. M., {\&} Ott, W. R. 1977, Applied Optics, 16, 367

Burris, D. L., Pilachowski, C. A., Armandroff, T. E., Sneden, C., Cowan, J. 
J., {\&} 

Roe, H. 2000, ApJ, 544, 302

Cayrel, R. et al. 2001, Nature, 409, 691

Cayrel, R. et al. 2004, A{\&}A, 416, 1117

Corliss, C. H., {\&} Bozman, W. R. 1962, 
\textit{Experimental Transition Probabilities for 
Spectral Lines of Seventy Elements}, U. S. Natl. Bur. Standards Monograph 53, 
(Washington: U. S. Government 
Printing Office)

Cowan, J. J., Burris, D. L., Sneden, C., McWilliam, A., {\&} Preston, G. W. 
1995, ApJ, 
439, L51

Cowan, J. J., {\&} Sneden, C. 2006, Nature, 440, 1151

Cowan, J. J., et al. 2002, ApJ,. 572, 861

Cowan, J. J., et al. 2005, ApJ, 627, 238

Cowan, J. J., Lawler, J. E., Sneden, C., Den Hartog, E. A., {\&} Collier, J. 
2006, To appear in Proc. NASA LAW ed. V. Kwong 

Danzmann, K., {\&} Kock, M. 1982, J. Opt. Soc. Am., 72, 1556

Delbouille, L, Roland, G., {\&} Neven, L. 1973, \textit{Photometric Atlas of the Solar Spectrum} \textit{from $\lambda $3000 to $\lambda $10000}, (Li\`{e}ge, Inst. D'Ap., 
Univ. de Li\`{e}ge)

Den Hartog, E. A., Herd, M. T., Lawler, J. E., Sneden, 
C., Cowan, J. J., {\&} Beers, T. C.
2005, Ap J. 619, 639

Den Hartog, E. A., Lawler, J. E., Sneden, C., {\&} Cowan, J. J. 2003, ApJS, 
148, 543

Den Hartog, E. A., Lawler, J. E., Sneden, C., {\&} Cowan, J. J. 2006, ApJS, 
in press

Den~Hartog, E. A., Wickliffe, M. E., {\&} Lawler, J. E. 2002, ApJS, 141, 255

Den~Hartog, E. A., Wiese, L. M. {\&} Lawler, J. E. 1999, J. Opt. Soc. Am. B, 
16, 2278

Duquette, D. W., Den Hartog, E. A., {\&} Lawler, J. E. 1986, J. Quant. 
Spect. Rad. Trans,
35, 281

Edl\'{e}n, B. 1953, J. Opt. Soc. Am., 43, 339

Gratton, R. G., {\&} Sneden, C. 1994, A{\&}Ap, 287, 927

Grevesse, N., {\&} Sauval, A. J. 1998, Space Sci. Rev., 85, 161

Grevesse, N., {\&} Sauval, A. J. 1999, A{\&}Ap, 347, 348

Grevesse, N., {\&} Sauval, A. J. 2002, Advances in Space Science Res., 30, 3

Grigoriev, I. S., {\&} Melikhov, E. Z. 1997, \textit{Handbook of Physical Quantities}, (Boca Raton,
CRC Press) p. 516

Guo, B., Ansbacher, W., Pinnington, E. H., Ji, Q., {\&} Berends, R. W. 1992, 
Phys. Rev. A,
46, 641

Hannaford, P, Lowe, R. M., Grevesse, N., Bi\'{e}mont, E., {\&} Whaling, W. 
1982,
ApJ, 261, 736

Hashiguchi, S., {\&} Hasikuni, M. 1985, J. Phys. Soc. Japan 54, 1290

Hill, V., et al. 2002, A{\&}A, 387, 560

Holweger, H., {\&} M\"{u}ller, E. A. 1974, Sol. Phys., 39, 19

Irwin, A. W. 1981, ApJS, 45, 621

Ivans, I. I., Simmerer, J., Sneden, C., Lawler, J. E., Cowan, J. J., 
Gallino, R., {\&} Bisterzo,
S. 2006, ApJ, 645, 613

Ivarsson, S., Litz\'{e}n, U., {\&} Wahlgren, G. M. 2001, Phys. Scr., 64, 455

K\"{a}ppeler, F., Beer, H., {\&} Wisshak, K. 1989, Rep. Prog. Phys., 52, 945

Klose, J. Z., Bridges, J. M., {\&} Ott, W. R. 1988, J. Res. Natl. Bur. 
Stand., 93, 21

Kono, A., {\&} Hattori, S. 1984, Phys. Rev. A, 29, 2981

Kratz, K. L., Farouqi, K., Pfeiffer, B., Truran, J. W., Sneden, C., {\&} 
Cowan, J. J. 2006, ApJ, submitted 

Kurucz, R. L. 1998, in \textit{Fundamental Stellar Properties: The Interaction between
Observation and Theory}, IAU Symp. 189, ed T. R. Bedding, A. J. Booth and J. 
Davis (Dordrecht: Kluwer), p. 217

Lawler, J. E., Bonvallet, G., {\&} Sneden, C. 2001a, ApJ, 556, 452

Lawler, J. E., Wickliffe, M. E., Den Hartog, E. A., {\&} Sneden, C. 2001b, 
ApJ, 563, 1075

Lawler, J. E., Wickliffe, M. E., Cowley, C. R., {\&} Sneden, C. 2001c, ApJS, 
137, 341

Lawler, J. E., Sneden, C., {\&} Cowan, J. J. 2004, ApJ, 604, 850

Lawler, J. E., Den Hartog, E. A., Sneden, C., {\&} Cowan, J. J. 2006, ApJS, 
162, 227 

Lodders, K. 2003, ApJ, 591, 1220

Lundqvist, M., Nilsson, H., Wahlgren, G. M., Lundberg, H., Xu, H. L., Jang, 
Z.-K., {\&}
Leckrone, D. S. 2006, A{\&}A, 450, 407

Malcheva, G., Blagoev, K., Mayo, R., Ortiz, M., Xu, H. L., Svanberg, S., 
Quinet, P., {\&}

Bi\'{e}mont, E. 2006, MNRAS, 367, 754

McWilliam, A., Preston, G. W., Sneden, C., {\&} Searle, L. 1995, AJ, 109, 
2757

Moore, C. E., Minnaert, M. G. J., {\&} Houtgast, J. 1966, 
\textit{The Solar Spectrum 2934 {\AA} to 
8770 {\AA}}, NBS Monograph 61 (Washington: U.S. G. P. O.)

Moore, C. E. 1971, \textit{Atomic Energy Levels}, Nat. Stand. Ref. Data 
Ser. Nat. Bur. Stand. 35,
V. III, (Washington: U. S. Government Printing Office), p. 147

O'Brien, S., Dababneh, S., Heil, M., K\"{a}ppeler, F., Plag, R., Reifarth, 
R., Gallino, R.,
Pignatari, M. 2003, Phys. Rev. C, 68, 035801-035807 

Palmeri, P., Quinet, P., Wyart, J.-F., {\&} Bi\'{e}mont, E. 2000, Physica 
Scripta, 61, 323

Pignatari, M., Gallino, R., Baldovin, C., Wiescher, M., Herig, F., Heger, 
A., Heil, M., {\&}
K\"{a}ppeler, F. 2006, To appear in the Proceedings of the International 
Symposium 
on Nuclear Astrophysics.

Ryan, S. G., Norris, J. E., {\&} Beers, T. C. 1996, ApJ, 471, 254

Schatz, H., Toenjes, R., Pfeiffer, B., Beers, T. C., Cowan, J. J., Hill, V., 
{\&} Kratz, K-L.
2002, ApJ, 579, 626 

Simmerer, J., Sneden, C., Cowan, J. J., Collier, J., Woolf, V.M., {\&} 
Lawler, J. E. 
2004, ApJ, 617, 1091 

Sneden, C. 1973, ApJ, 184, 839

Sneden, C. et al. 2003, ApJ, 591, 936

Sneden, C., McWilliam, A., Preston, G. W., Cowan, J. J., Burris, D. L., {\&} 
Armosky, B. 
J. 1996, ApJ,  467, 819

Volz, U., {\&} Schmoranzer, H. 1998, in AIP Conf. Proc. 434, 
\textit{Atomic and Molecular Data 
and Their Applications}, ed. P. J. Mohr {\&} W. L. Wiese (Woodbury, NY:AIP), p. 67

Weiss, A. W. 1995, Phys. Rev. A, 51, 1067

Westin, J., Sneden, C., Gustafsson, B., {\&} Cowan, J.J. 2000, ApJ, 530, 783

Whaling, W., Carle, M. T., {\&} Pitt, M. L. 1993, J. Quant. Spectrosc. Rad. 
Transfer 50, 7

Wickliffe, M. E., Lawler, J. E., {\&} Nave, G. 2000, J. Quant. Spec. Rad. 
Transfer,
66, 363

Winckler, N., Dababneh, S., Heil, M., K\"{a}ppeler, F., Gallino, R., {\&} 
Pignatari, M. 2006,
ApJ, 647, 685

Wyart, J-F. {\&} Blaise J. 1990, Phys. Scr., 42, 209

Yan, Z-C, Tambasco, M., {\&} Drake, G. W. F. 1998, Phys. Rev. A, 57, 1652

\newpage
\begin{center}
FIGURE CAPTIONS
\end{center}
Figure 1. Ratio of the Lund University (Lund) branching fraction from 
(Lundqvist et al. 2006) to the University of Wisconsin (UW) branching 
fraction from this work as a function of wavelength.

Figure 2. Ratio of the Lund University branching fraction (Lundqvist et al. 
2006) to the University of Wisconsin (UW) branching fraction from this work 
as a function of the UW branching fraction.

Figure 3. Delta log(gf) = log(gf)$_{Lund}$ -- log(gf)$_{UW}$ plotted as a 
function of log(gf)$_{UW}$.

Figure 4. Relative strength factors, log($\varepsilon $\textit{gf}) -- $\theta \chi 
$, ($\equiv $STR) for Gd \textsc{ii} and Hf \textsc{ii} lines, 
plotted as functions of wavelength. These factors are discussed in the text 
in {\S} 4.1. Solar abundances ($\varepsilon _{Sun})$ and an approximate 
inverse temperature ($\theta $ = 1.0) have been assumed in forming these 
factors. Dotted lines indicate the approximate value of STR for lines near 
the detection limit in the Sun, and dashed lines indicate minimum STR levels 
for lines that are strong in the Sun. The left-hand panel is identical to 
the right-hand panel of Figure 4 in DLSC06. The right-hand panel is 
generated with data from the present work. In this panel we have shown with 
circled dots the 12 Hf \textsc{ii} lines finally employed in the abundance 
analyses.

Figure 5. Spectra of the four Hf \textsc{ii} lines used in the solar 
photospheric analysis. The filled circles represent the center-of-disk 
spectra of Delbouille et al. (1973), but for clarity in the figure we have 
only shown points spaced every 0.01 {\AA} instead of the original 0.002 
{\AA}. The four lines shown in each panel represent the synthetic spectra 
for which the hafnium abundance has been varied. The synthetic spectrum with 
weak-to-absent absorption at the Hf \textsc{ii} wavelength was computed 
without any hafnium contribution. The synthetic spectrum that nearly traces 
the observed one was computed with the log $\varepsilon $(Hf) value derived 
for that transition (see Table 5), and the other two syntheses were done 
with hafnium abundances decreased and increased by a factor of two from the 
best-fit value. The labeled tick marks in each panel are put at 1 {\AA} 
intervals, and the unlabeled tick marks are spaced at 0.1 {\AA}. 

Figure 6. Abundance comparisons of log $\varepsilon $(La/Eu) versus 
[Fe/H] and [Eu/H] for a group of 10 metal-poor $r-$process rich stars. The 
dotted lines define the range of the solar system $r-$process-only values based 
upon the published deconvolution of the solar system abundances (Simmerer et 
al. 2004, Cowan et al. 2006), the solid line is the mean ratio of the 10 
halo stars in our sample, and the dashed line is the total solar system 
ratio based upon the stellar value for the $r-$process (see text in {\S} 5.1 for 
details).

Figure 7. Abundance comparisons of log $\varepsilon $(Hf/Eu) versus 
[Fe/H] and [Eu/H] for a group of 10 metal-poor r-process rich stars. The 
lines are similar to those in Figure 6.

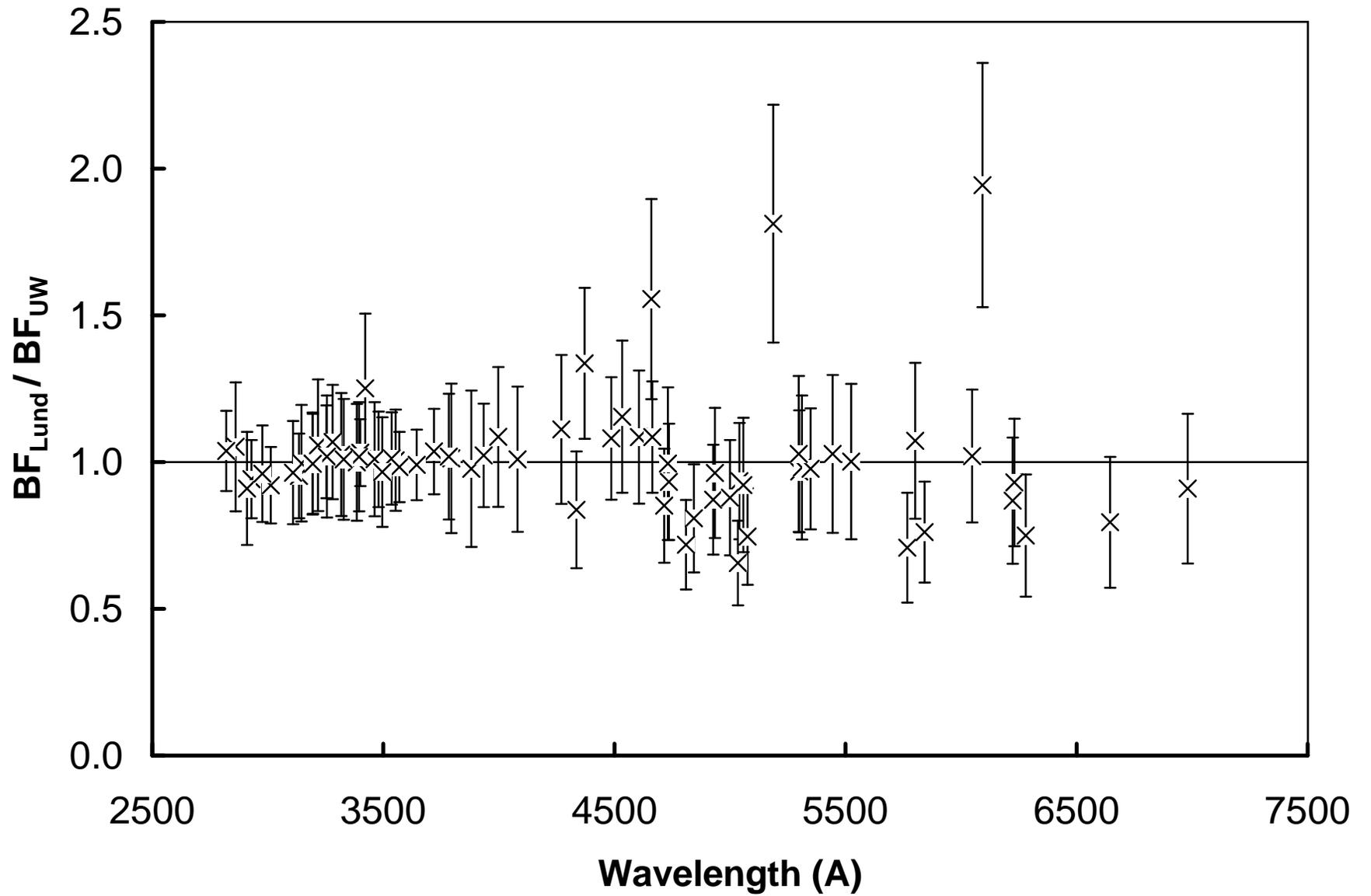
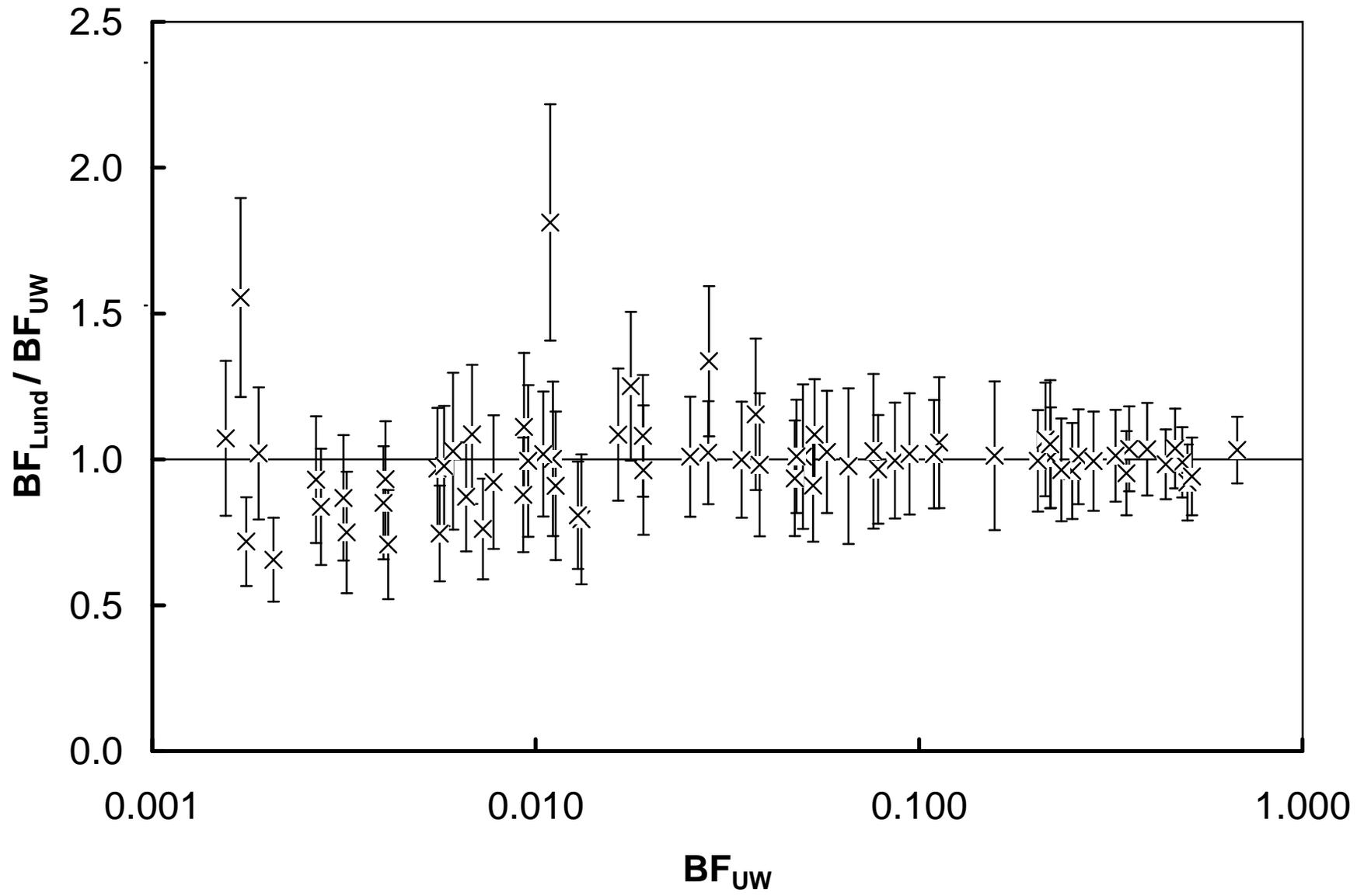
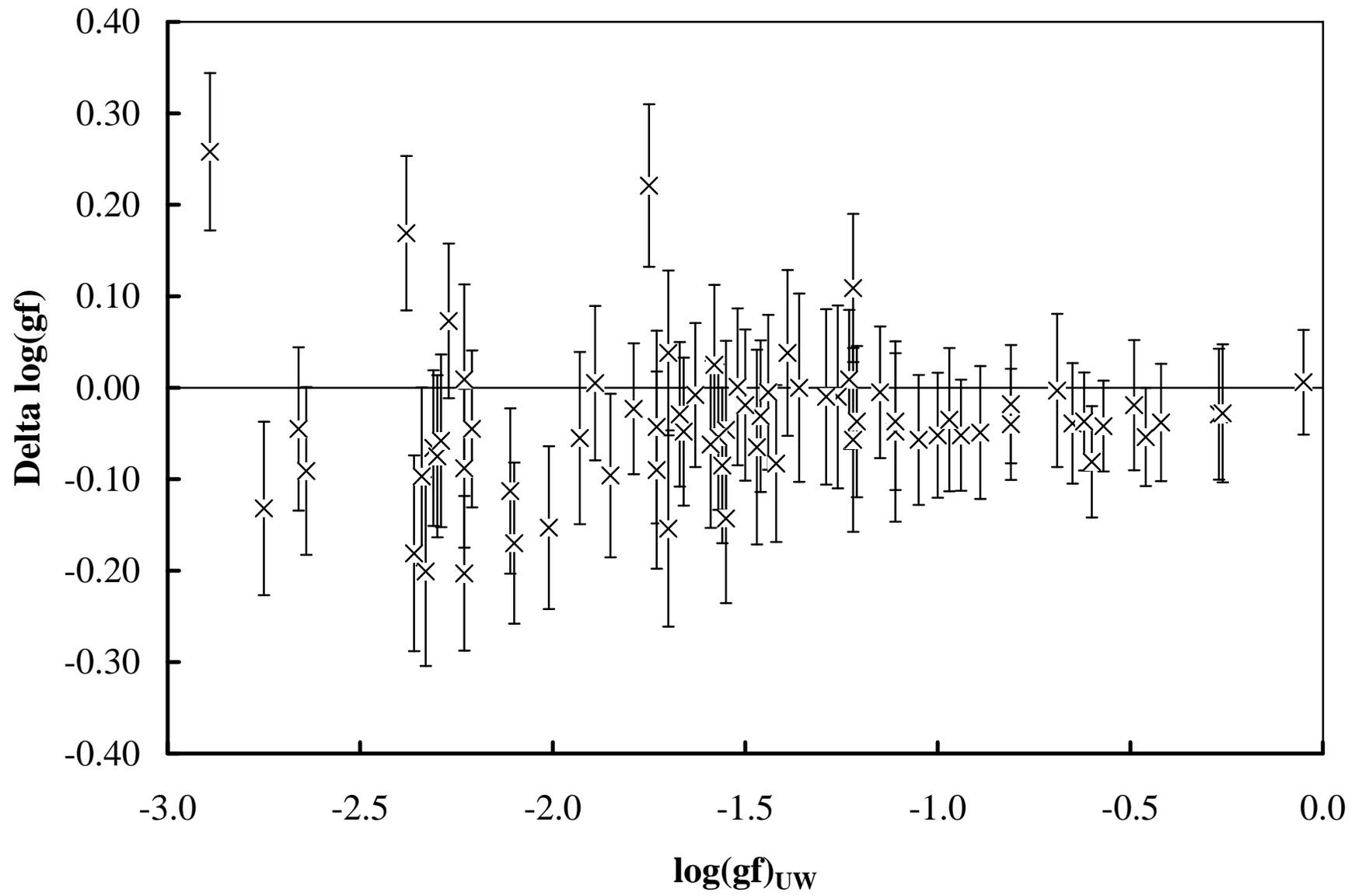
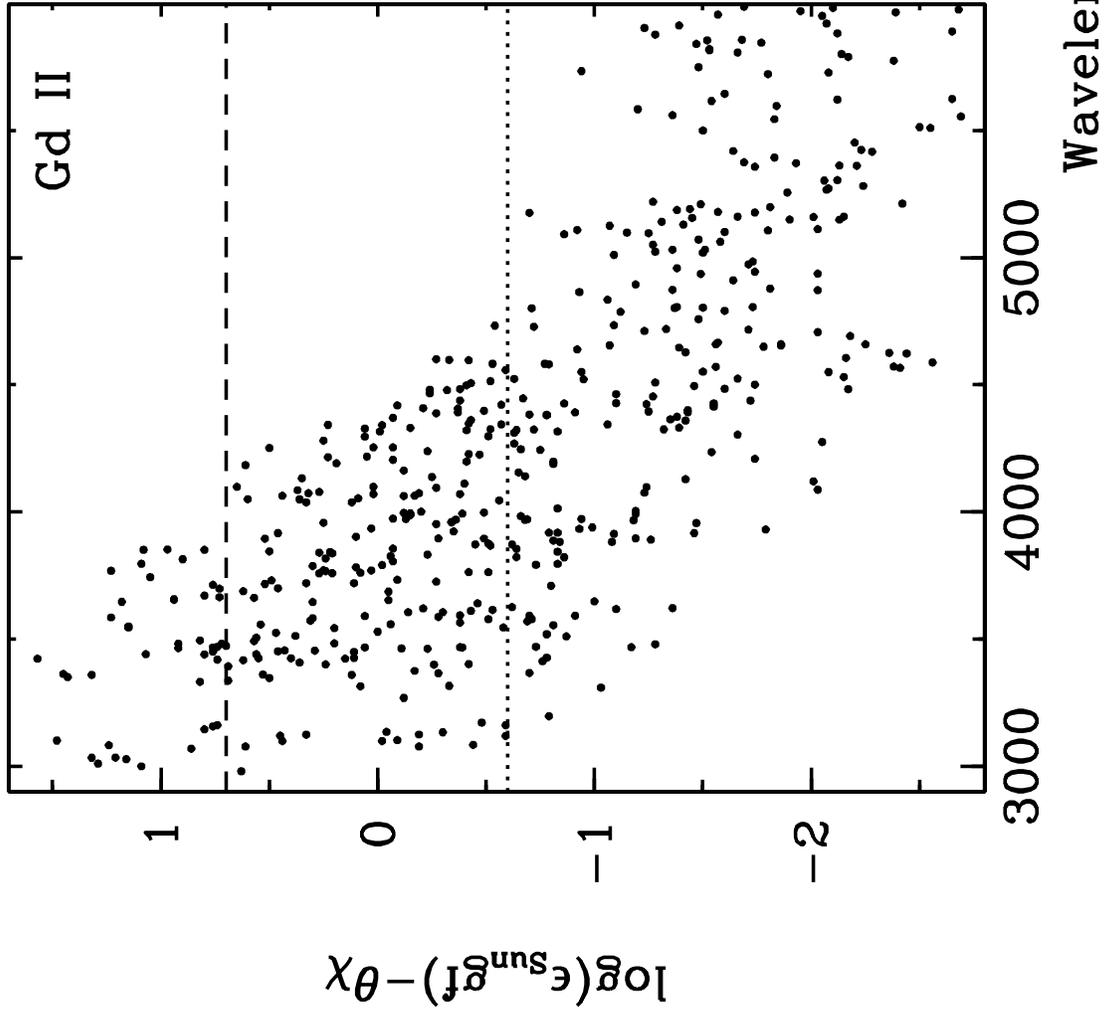
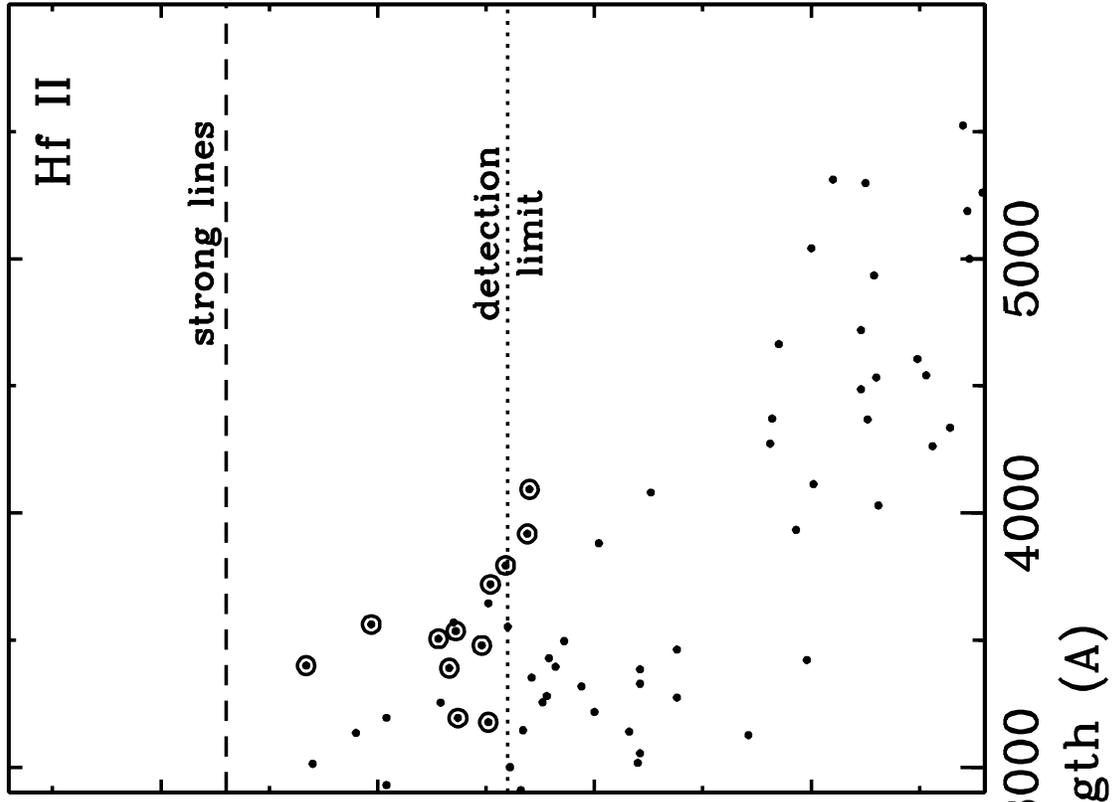
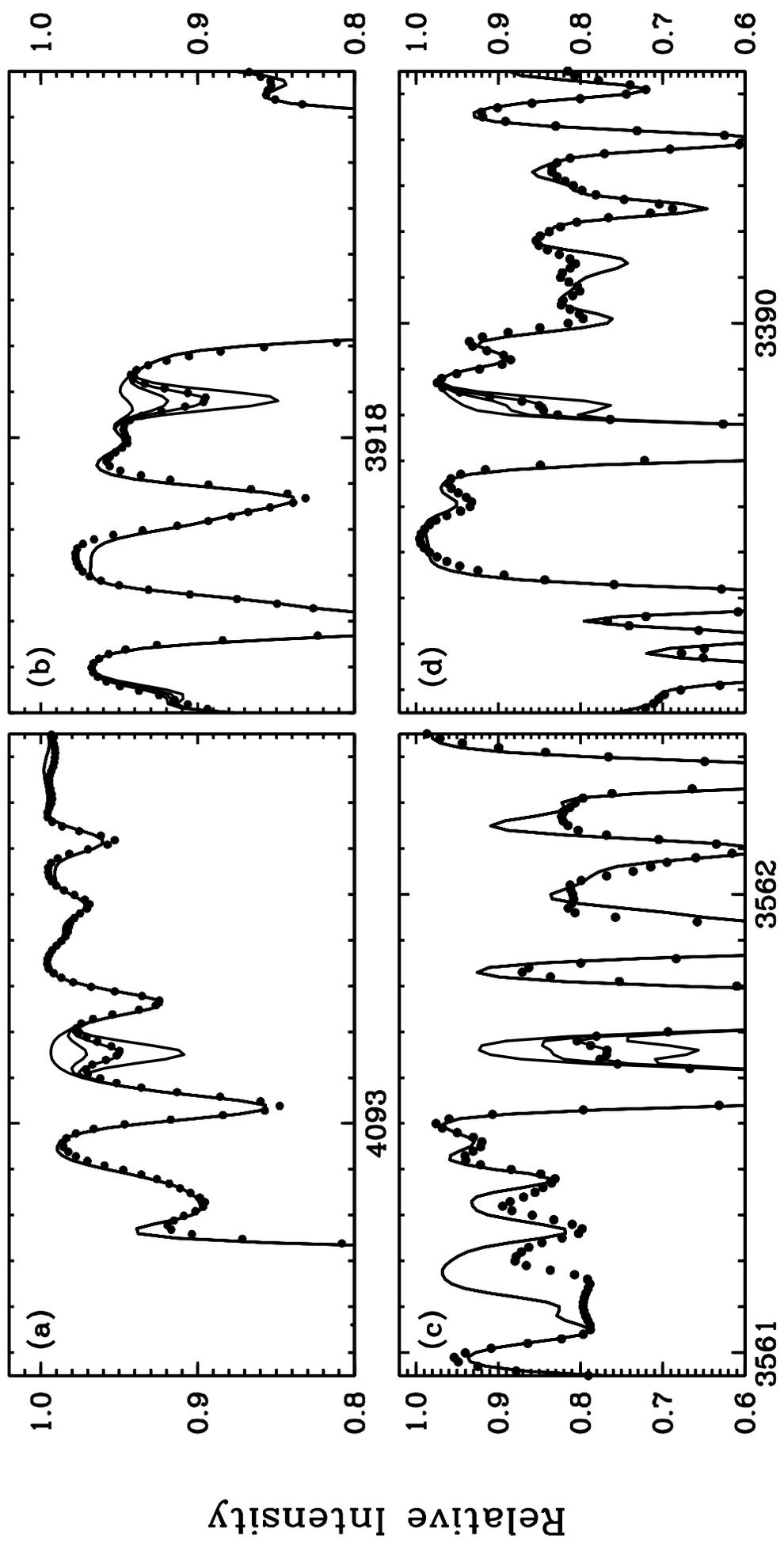
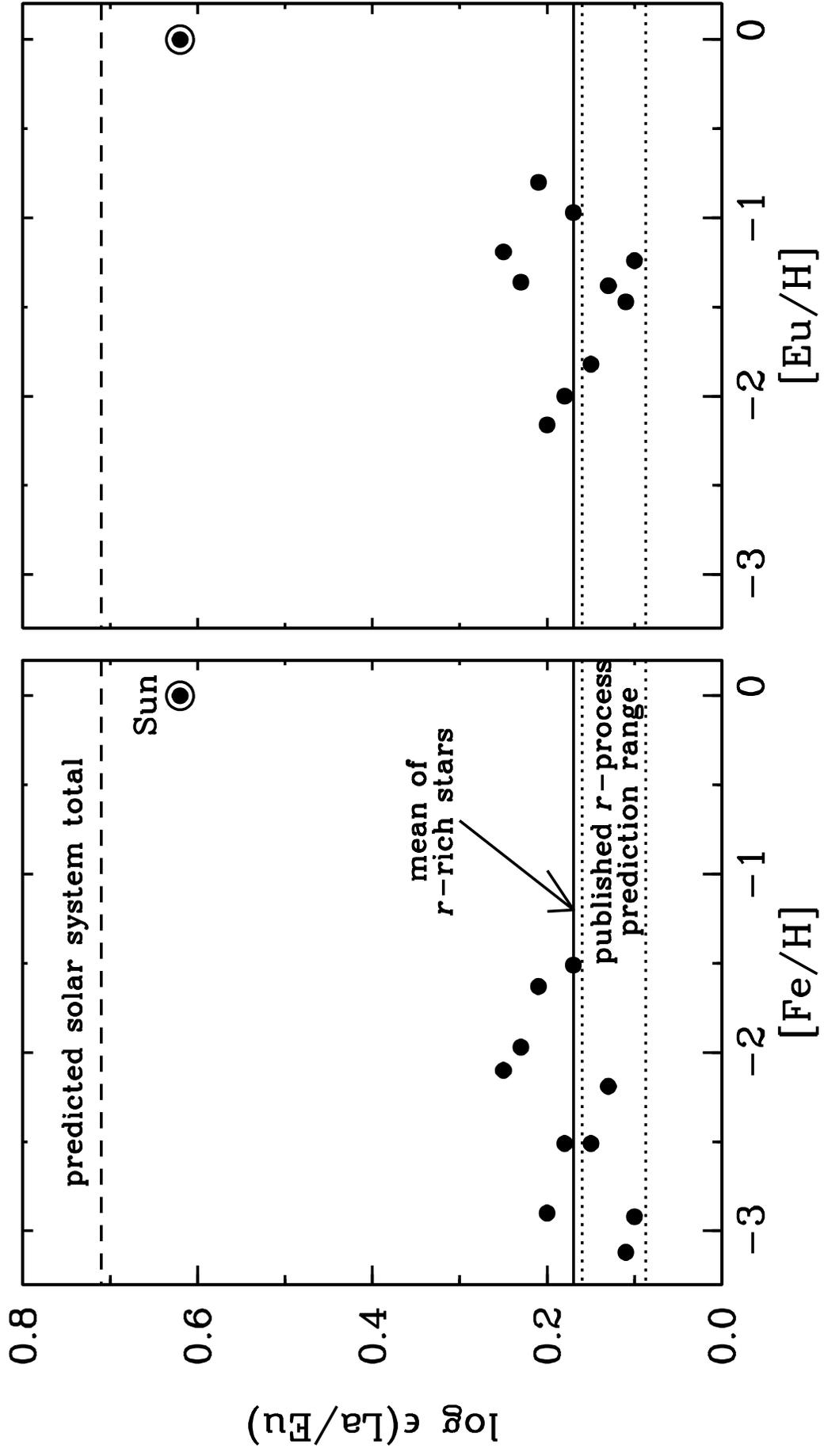
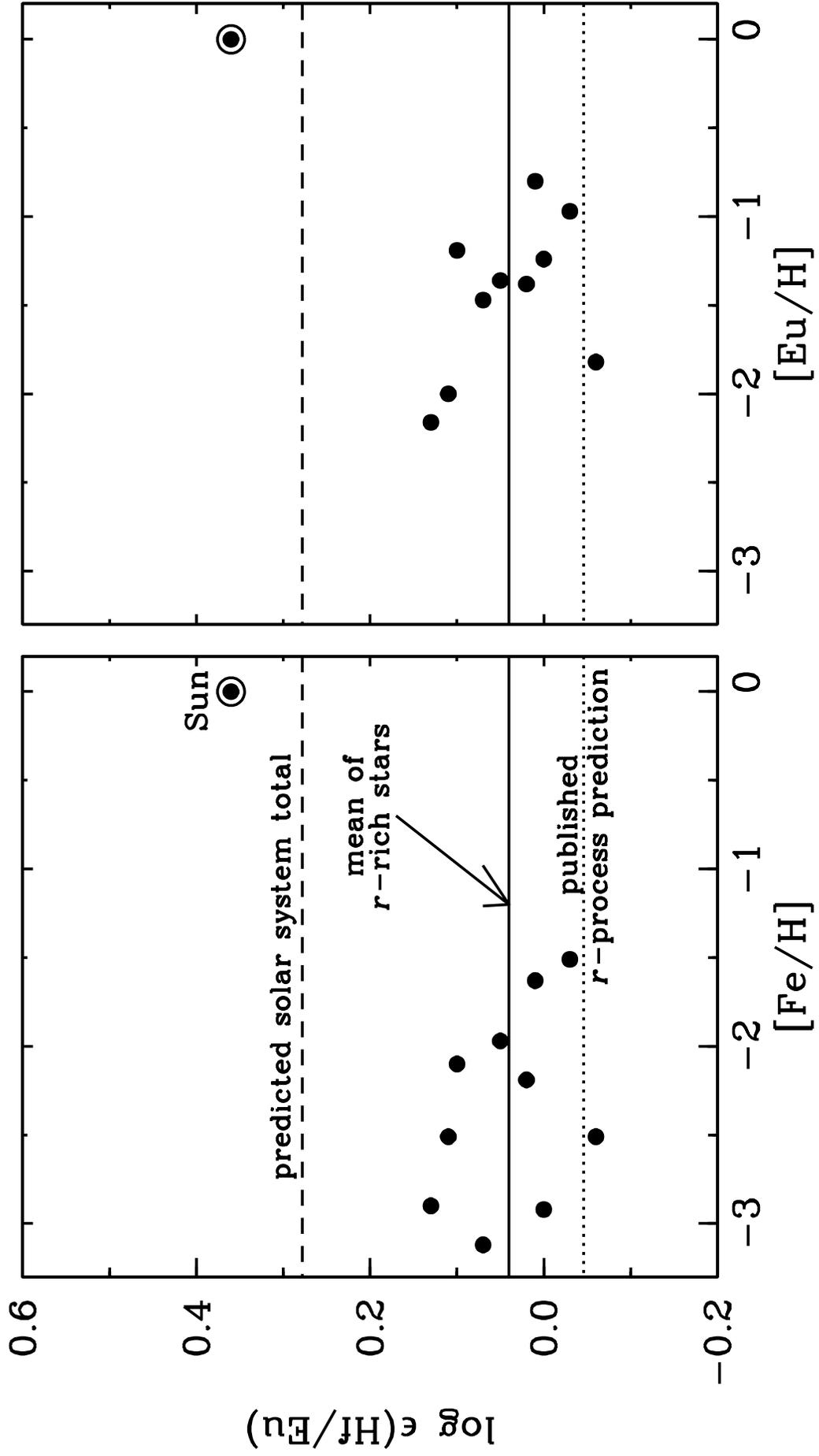
\end{document}


\begin{deluxetable}{cccccllccccc} 
\tabletypesize{\scriptsize}
\rotate
\tablenum{2}
\tablecolumns{12} 
\tablewidth{0pc} 
\tablecaption{Fourier transform spectra of Hf hollow cathode discharge lamps used 
in this study. All were recorded using the 1 m FTS on the McMath telescope 
at the National Solar Observatory, Kitt Peak, AZ.} 
\tablehead{
\colhead{Index}& 
\colhead{Date}& 
\colhead{Serial}& 
\colhead{Buffer}& 
\colhead{Lamp}& 
\colhead{Wavenumber}& 
\colhead{Limit of}& 
\colhead{Coadds}& 
\colhead{Beam}& 
\colhead{Filter}& 
\colhead{Detector}{\hspace*{-0.20in}$^{a}$}& 
\colhead{Calibration}$^{b}$
\\
& 
& 
\colhead{Number}& 
\colhead{Gas}& 
\colhead{Current}& 
\colhead{Range}& 
\colhead{Resolution}& 
& 
\colhead{Splitter}& 
& 
& 
\\
& 
& 
& 
& 
\colhead{(mA)}& 
\colhead{(cm}$^{-1}$)& 
\colhead{(cm}$^{-1}$)& 
& 
& 
& 
& 
}

\startdata
1& 
1984 Mar. 20& 
10& 
Ar& 
93& 
0.00 - 36947& 
0.053& 
8& 
UV& 
& 
Midrange Si Diode& 
Ar I {\&} II \par  \\
2& 
1984 Mar. 21& 
2& 
Ar& 
43& 
14432 - 36081& 
0.053& 
8& 
UV& 
CuSO$_{4}$& 
Midrange Si Diode& 
Ar I {\&} II \\
3& 
1984 Mar. 21& 
3& 
Ar& 
169& 
14432 - 36081& 
0.053& 
8& 
UV& 
CuSO$_{4}$& 
Midrange Si Diode& 
Ar I {\&} II \\
4& 
1984 Nov. 1& 
1& 
Ar& 
160& 
5922 - 44951& 
0.064& 
8& 
UV& 
CS 9-54& 
Midrange Si Diode& 
Ar I {\&} II \\
5& 
1984 Nov. 1& 
2& 
Ar& 
100& 
5922 - 44951& 
0.064& 
8& 
UV& 
CS 9-54& 
Midrange Si Diode& 
Ar I {\&} II \\
6& 
1987 Jan. 14& 
9& 
Ar& 
500& 
8218 - 26091& 
0.050& 
8& 
Visible& 
& 
S. B. Si Diode& 
Ar I {\&} II,  W Strip Lamp \\
7& 
1984 Mar. 20& 
11& 
Ar& 
93& 
0.00 - 22831& 
0.033& 
8& 
UV& 
& 
Midrange Si Diode& 
Ar I {\&} II \par  \\
8& 
1984 Mar. 21& 
1& 
Ar& 
42& 
7491 - 22028& 
0.033& 
8& 
UV& 
GG 495& 
Midrange Si Diode& 
Ar I {\&} II \\
9& 
1984 Mar. 21& 
4& 
Ar& 
42& 
7491 - 22028& 
0.033& 
8& 
UV& 
GG 495& 
Midrange Si Diode& 
Ar I {\&} II \\
10& 
1984 Nov. 1& 
3& 
Ar& 
100& 
3170 -- 15058& 
0.022& 
8& 
UV& 
RG 715& 
InSb& 
Ar I {\&} II \\
11& 
1984 Nov. 1& 
4& 
Ar& 
180& 
3170 -- 15058& 
0.022& 
8& 
UV& 
RG 715& 
InSb& 
Ar I {\&} II \\
12& 
1987 Jan. 14& 
10& 
Ne& 
330& 
8218 - 26091& 
0.050& 
8& 
Visible& 
& 
S. B. Si Diode& 
W Strip Lamp \\
13& 
1987 Jan. 14& 
11& 
Ne& 
370& 
3488 - 15017& 
0.029& 
8& 
Visible& 
& 
InSb& 
W Strip Lamp
\enddata

\tablenotetext{a}{Detectors types include the Midrange Si photodiode, Super Blue (S. B.) 
Si photodiode, and InSb dectector for the Near IR.}
\tablenotetext{b}{Relative radiometric calibrations were based on selected sets of Ar I 
and Ar II lines and on a Tungsten (W) Strip Lamp calibrated as a secondary 
radiance standard.}

\end{deluxetable}